\documentstyle[prd,aps]{revtex}

\newcommand{\bea}{\begin{eqnarray}}
\newcommand{\eea}{\end{eqnarray}}

\begin{document}

\draft
\twocolumn[\hsize\textwidth\columnwidth\hsize\csname
@twocolumnfalse\endcsname

\title{Second-order perturbations of a zero-pressure cosmological medium: \\
       Proofs of the relativistic-Newtonian correspondence}
\author{Jai-chan Hwang${}^{(a)}$ and Hyerim Noh${}^{(b)}$}
\address{${}^{(a)}$ Department of Astronomy and Atmospheric Sciences,
                    Kyungpook National University, Taegu, Korea \\
         ${}^{(b)}$ Korea Astronomy and Space Science Institute,
                    Daejon, Korea
         }
\date{\today}
\maketitle

\begin{abstract}

The dynamic world model and its linear perturbations were first
studied in Einstein's gravity. In the system without pressure the
relativistic equations coincide exactly with the later known ones in
Newton's gravity. Here we {\it prove} that, except for the
gravitational wave contribution, even to the second-order
perturbations, equations for the relativistic irrotational
zero-pressure fluid in a flat Friedmann background {\it coincide}
exactly with the previously known Newtonian equations. Thus, to the
second order, we correctly identify the relativistic density and
velocity perturbation variables, and we {\it expand} the range of
applicability of the Newtonian medium without pressure to {\it all}
cosmological scales including the super-horizon scale. In the
relativistic analyses, however, we do {\it not} have a relativistic
variable which corresponds to the Newtonian potential to the second
order. Mixed usage of different gauge conditions is useful to make
such proofs and to examine the result with perspective. We also
present the gravitational wave equation to the second order. Since
our correspondence includes the cosmological constant, our results
are relevant to currently favoured cosmology. Our result has an
important practical implication that one can use the large-scale
Newtonian numerical simulation more reliably even as the simulation
scale approaches near horizon.

\end{abstract}

\vskip2pc]
%
%
\section{Introduction}

Despite its algebraic and conceptual complexity in Einstein's
gravity the evolving world model and its linear structures were
first studied based on Einstein's gravity in the classic works by
Friedmann in 1922 \cite{Friedmann-1922} and Lifshitz in 1946
\cite{Lifshitz-1946}, respectively. In an interesting sequence, the
much simpler and, in hindsight, more intuitive Newtonian studies
followed later by Milne in 1934 \cite{Milne-1934} and Bonnor in 1957
\cite{Bonnor-1957}, respectively. In the case without pressure the
Newtonian results {\it coincide} exactly with the previously derived
relativistic ones for both the background world model and its
first-order (linear) perturbations. The case with pressure {\it
cannot} be handled in the Newtonian context despite several failed
attempts in the literature to simulate it especially for the
perturbation. The situation is still well described by Sachs and
Wolfe in 1967 \cite{SW-1967}:
      ``When these modified equations were perturbed to first order,
        their solutions did not agree with the relativistic results,
        even qualitatively.''
In this work, we will show an additional continuation of
relativistic-Newtonian correspondences in the zero-pressure medium
by {\it proving} that the relativistic second-order scalar-type
perturbations are described by the same equations known in Newton's
theory. That is, the Newtonian equations {\it coincide} exactly with
the relativistic ones even to the second order in perturbations.

In the relativistic perturbations, due to the covariance of field
equation we have freedom to fix the spacetime coordinate system by
choosing some of the metric or energy-momentum variables at our
disposal: this is often called the gauge choice. The original study
of Lifshitz started by choosing the synchronous gauge which is still
quite popular in the literature. Other gauge conditions were
discovered later \cite{Harrison-Nariai,Bardeen-1980}. It is an
ironic situation that except for the widely used synchronous gauge
condition, each of other gauge conditions fixes the gauge freedom
completely. Thus, each has its own unique corresponding gauge
invariant combination. Notice some common algebraic errors (not in
Lifshitz's work though) widespread in the literature including many
textbooks due to the incomplete gauge fixing nature of the
synchronous gauge, see \cite{GRG}.

Although infinitely many gauge conditions are available, it has been
common in the literature to fix gauge conditions from the beginning.
The importance of using different gauge conditions for different
variables and the gauge invariance of such variables were shown by
Bardeen in 1980 \cite{Bardeen-1980}. Bardeen's work also showed the
importance of having access to many different gauge conditions which
become apparent in his work in 1988 \cite{Bardeen-1988}. In this
work, the importance of having different variables evaluated in
different gauges (all correspond to unique gauge-invariant
combinations) will become clear as we extend Bardeen's approach to
the second-order perturbations.

Recently, we have presented a second-order perturbation formulation
of the Friedmann world model considering quite general situations
\cite{NL}. We have resolved the gauge issue, identifying the
variables to use in fixing the gauges and constructing
gauge-invariant combinations, which can be easily extended even to
the higher order. The basic equations are presented without fixing
the temporal gauge condition thus allowing us to choose or try many
available gauge conditions later depending on the situation: we call
this a gauge-ready approach, see Eqs.\
(\ref{scalar-0})-(\ref{scalar-6}) below. The Newtonian
correspondence to the linear order was made by properly arranging
the equations using various gauge-invariant variables in
\cite{Bardeen-1980,H-MDE-1994,HN-Newtonian-1999}. Extending such
correspondences to the second order is our task in this work. We set
$c \equiv 1$.

%
%
\section{Basic equations}

We consider a {\it scalar-type} perturbation in the {\it flat}
Friedmann background. We will consider the presence of tensor-type
perturbation (gravitational waves) in \S \ref{sec:Proof2}. The
vector-type perturbation (rotation) is not important because it
always decays in the expanding phase even to the second order, see
\S VII.E of \cite{NL}. Our reason for considering the flat
background will be explained below Eq.\ (\ref{pert}). As the metric
we take \bea
   ds^2
   &=& - a^2 \left( 1 + 2 \alpha \right) d \eta^2
       - 2 a^2 \beta_{,\alpha} d \eta d x^\alpha
   \nonumber \\
   & &
       + a^2 \left[ g^{(3)}_{\alpha\beta} \left( 1 + 2 \varphi \right)
       + 2 \gamma_{,\alpha|\beta} \right] d x^\alpha d x^\beta,
   \label{metric}
\eea which follows from our convention in Eqs.\ (49), (175), and
(178) of \cite{NL}. Here, $a(t)$ is the scale factor, and $\alpha$,
$\beta$, $\gamma$ and $\varphi$ are spacetime dependent
perturbed-order variables; we take Bardeen's metric convention in
\cite{Bardeen-1988} extended to the second order. A vertical bar
indicates a covariant derivative based on $g^{(3)}_{\alpha\beta}$
which becomes $\delta_{\alpha\beta}$ if we take Cartesian
coordinates in the flat Friedmann background. By taking $\gamma
\equiv 0$, which we call the spatial $C$-gauge, the spatial gauge
mode is removed completely, thus all the remaining variables we are
using are spatially gauge-invariant to the second order, see \S
VI.B.2 of \cite{NL}. In the following we will take $\gamma \equiv 0$
as the spatial gauge condition and use $\chi \equiv a \beta + a^2
\dot \gamma$ which becomes $\chi = a \beta$.

As the energy-momentum tensor we take \bea
   \tilde T^0_0
   &=& - \mu - \delta \mu
       + {1 \over a} \mu \chi^{,\alpha} v_{,\alpha},
   \nonumber \\
   \tilde T^0_\alpha
   &=& - \mu \left( 1 - \alpha \right) v_{,\alpha},
   \nonumber \\
   \tilde T^\alpha_\beta
   &=& \delta p \delta^\alpha_\beta
       + {1 \over a^2} \left( \Pi^{,\alpha}_{\;\;\; |\beta}
       - {1 \over 3} \delta^\alpha_\beta \Delta \Pi \right)
       - {1 \over a} \mu \chi^{,\alpha} v_{,\beta},
   \label{Tab-pert-normal}
\eea which follows from our convention in Eqs.\ (84), (175), and
(178) of \cite{NL}; tildes indicate the covariant quantities. Here,
$\mu$ is the background energy density, and $\delta \mu$, $\delta
p$, $\Pi$ and $v$ are the perturbed order energy-density, isotropic
pressure, anisotropic pressure, and the flux, respectively, all
based on the normal frame vector $\tilde n_a$ with $\tilde n_\alpha
\equiv 0$. Although we are considering a zero-pressure system (thus,
$p= 0$ and $\delta p = 0 = \Pi$ to the linear order), it is
essential to keep the perturbed pressure terms $\delta p$ and $\Pi$
because these do not necessarily vanish to the second order in
perturbation depending on the coordinate (gauge) condition we
choose. This is because in \cite{NL} we have evaluated the fluid
quantities based on the normal-frame $\tilde n_a$; we will elaborate
this point in \S \ref{sec:Pressures}.

To the background order we have the Friedmann equation
\cite{Friedmann-1922,Milne-1934,MM-1934}
\bea
   & & H^2 = {8 \pi G \over 3} \mu
       - {{\rm const.} \over a^2}
       + {\Lambda \over 3},
   \label{BG}
\eea with the energy (mass) density $\mu$ $(\varrho) \propto
a^{-3}$; $\Lambda$ is the cosmological constant. To the linear-order
perturbations we have a second-order differential equation
originally derived by Lifshitz \cite{Lifshitz-1946,Bonnor-1957} \bea
   & & \ddot \delta + 2 H \dot \delta - 4 \pi G \mu \delta
       = 0.
   \label{pert}
\eea An overdot indicates a time derivative based on $t$ ($dt \equiv
a d \eta$) and $H \equiv {\dot a \over a}$. The variable $a(t)$ is
the scale factor, and $\delta \equiv {\delta \mu \over \mu} =
{\delta \varrho \over \varrho}$ with $\mu$ $(\varrho)$ and $\delta
\mu$ $(\delta \varrho)$ the background and perturbed parts,
respectively, of the energy (mass) density field. The ``{\rm
const.}'' part is interpreted as the spatial curvature in Einstein's
gravity and the total energy in the Newton's gravity \cite{MM-1934}.
Equation (\ref{pert}) is valid even in the presence of the
cosmological constant $\Lambda$ as well as the background curvature.
We will include $\Lambda$ term in the following. In the relativistic
context Eq.\ (\ref{pert}) can be derived in the comoving gauge
condition; the original derivation by Lifshitz is based on the
synchronous gauge, and in the zero-pressure medium to the linear
order the synchronous gauge coincides with the comoving gauge:
further discussion about this point will be made in \S
\ref{sec:Proof}. Although Eq.\ (\ref{pert}) is also valid with
general spatial curvature the relativistic-Newtonian correspondence
is somewhat {\it ambiguous} in the case with curvature, for details
see \S 3 of \cite{HN-Newtonian-1999}. Thus, we consider the {\it
flat} background only.

The perturbed parts of equations to the second order are presented
in Eqs.\ (195)-(201) of \cite{NL}. In a flat background with
vanishing background pressure we have \bea
   \kappa - 3 H \alpha + 3 \dot \varphi + {\Delta \over a^2} \chi
   &=& N_0,
   \label{scalar-0} \\
   4 \pi G \delta \mu + H \kappa + {\Delta \over a^2} \varphi
   &=& {1\over 4} N_1,
   \label{scalar-1} \\
   \kappa + {\Delta \over a^2} \chi - 12 \pi G \mu a v
   &=& N_2^{({\rm s})},
   \label{scalar-2} \\
   \dot \kappa + 2 H \kappa
       - 4 \pi G \left( \delta \mu + 3 \delta p \right)
       + \left( 3 \dot H + {\Delta \over a^2} \right) \alpha
   &=& N_3,
   \label{scalar-3} \\
   \dot \chi + H \chi - \varphi - \alpha - 8 \pi G \Pi
   &=& N_4^{({\rm s})},
   \label{scalar-4} \\
   \delta \dot \mu + 3 H \left( \delta \mu + \delta p \right)
       - \mu \left( \kappa - 3 H \alpha
       + {\Delta \over a} v \right)
   &=& N_5,
   \label{scalar-5} \\
   {\left( a^4 \mu v \right)^\cdot \over a^4 \mu}
       - {1 \over a} \alpha
       - {1 \over a \mu} \left( \delta p
       + {2 \over 3} {\Delta \over a^2} \Pi \right)
   &=& N_6^{({\rm s})},
   \nonumber \\
   \label{scalar-6}
\eea where the pure quadratic-order terms, $N_i$, can be read from
Eqs.\ (99)-(105) in \cite{NL}. $\Delta$ is a Laplacian operator.
Equation (\ref{scalar-0}) is a definition of $\kappa$, Eqs.\
(\ref{scalar-1})-(\ref{scalar-4}) follow from $\tilde G^0_0$,
$\tilde G^0_\alpha$, $\tilde G^\alpha_\alpha - \tilde G^0_0$ and
$\tilde G^\alpha_\beta - {1 \over 3} \delta^\alpha_\beta \tilde
G^\gamma_\gamma$ components of Einstein's equation, respectively,
and Eqs.\ (\ref{scalar-5}), (\ref{scalar-6}) follow from $\tilde
T^b_{0;b} = 0$ and $\tilde T^b_{\alpha;b} = 0$, respectively. To the
linear order these set of equations without fixing the temporal
gauge was presented by Bardeen in \cite{Bardeen-1988}. All our
equations {\it include} the cosmological constant in the background.
These equations are presented without fixing the temporal gauge
condition and using only the spatially gauge-invariant variables
even to the second order; our choice of the spatial $C$-gauge
($\gamma \equiv 0$) guarantees such invariances of the remaining
variables, see \S VI.B of \cite{NL}. As the proper temporal gauge
condition we can choose any of the following: $\alpha \equiv 0$ (the
synchronous gauge), $\chi \equiv 0$ (the zero-shear gauge), $\delta
\equiv 0$ (the uniform-density gauge), $\kappa \equiv 0$ (the
uniform-expansion gauge), $v \equiv 0$ (the comoving gauge),
$\varphi \equiv 0$ (the uniform-curvature gauge), etc. Except for
the synchronous gauge, each of the other temporal gauge conditions
completely removes the temporal gauge mode. We can also take linear
combinations of the above conditions, and choose different gauge
conditions to different order, see \S VI.C.2 of \cite{NL}. Thus, we
have infinite number of different temporal gauge choices available
to each order in perturbations.

{}From Eqs.\ (\ref{scalar-0})-(\ref{scalar-6}) we can derive the
following set of equations expressed using gauge-invariant variables
\bea
   & & \alpha_v =
       - {1 \over 2} v_\chi^{\;\;,\alpha} v_{\chi,\alpha}
       - {1 \over \mu} \left( \delta p_v
       + {2 \over 3} {\Delta \over a^2} \Pi_v \right),
   \label{alpha_v-eq} \\
   & & \dot \delta_v - \kappa_v =
       {1 \over a} \left( \delta_v v_\chi^{\;\;,\alpha} \right)_{,\alpha}
       - 3 {H \over \mu} \delta p_v,
   \label{delta_v-eq} \\
   & & \dot \kappa_v + 2 H \kappa_v - 4 \pi G \mu \delta_v =
       {\Delta \over 2 a^2}
       \left( v_\chi^{\;\;,\alpha} v_{\chi,\alpha} \right)
       + 12 \pi G \delta p_v,
   \nonumber \\
   \label{kappa_v-eq} \\
   & & \kappa_v - {\Delta \over a} v_\chi =
       {1 \over a} \left( v_\chi \Delta \varphi_\chi
       - 2 \varphi_\chi \Delta v_\chi
       + \varphi_\chi^{\;\;,\alpha} v_{\chi,\alpha} \right)
   \nonumber \\
   & & \quad
       + {5 \over 2} H \left( 2 v_\chi \Delta v_\chi
       + v_\chi^{\;\;,\alpha} v_{\chi,\alpha} \right)
   \nonumber \\
   & & \quad
       - {1 \over a} \nabla^\alpha \left( \delta_v v_{\chi,\alpha} \right)
       - {3 \over a} \Delta^{-1} \nabla^\alpha \left(
       v_{\chi,\alpha} \Delta \varphi_\chi \right),
   \label{kappa_v-v_chi-eq} \\
   & & \alpha_\chi + \varphi_\chi =
       \varphi_\chi^2
       - \Delta^{-1} \left( \varphi_\chi \Delta \varphi_\chi \right)
   \nonumber \\
   & & \quad
       + 3 \Delta^{-2} \nabla^\alpha \nabla^\beta
       \left( \varphi_\chi \varphi_{\chi,\alpha\beta} \right)
       - 8 \pi G \Pi_\chi,
   \label{alpha_chi-eq} \\
   & & 4 \pi G \mu \delta_v + {\Delta \over a^2} \varphi_\chi =
       {1 \over 2} \dot H \Delta v_\chi^2
       - 3 a H \dot H \Delta^{-1} \nabla^\alpha
       \left( \delta_{v,\alpha} v_\chi \right)
   \nonumber \\
   & & \quad
       + {1 \over a^2} \left( 4 \varphi_\chi \Delta \varphi_\chi
       + {3 \over 2} \varphi_\chi^{\;\;,\alpha} \varphi_{\chi,\alpha} \right),
   \label{Poisson-eq} \\
   & & \dot v_\chi + H v_\chi - {1 \over a} \alpha_\chi =
       - {3 \over 2} a \dot H v_\chi^2
       + 3 H \varphi_\chi v_\chi
       - {1 \over 2a} \varphi_\chi^2
   \nonumber \\
   & & \quad
       - {1 \over a} \Delta^{-1} \nabla^\alpha
       \left( \delta_v \varphi_{\chi,\alpha} \right)
       + {1 \over a \mu} \left( \delta p_\chi
       + {2 \over 3} {\Delta \over a^2} \Pi_\chi \right),
   \label{v_chi-eq} \\
   & & \dot \varphi_\chi - H \alpha_\chi + 4 \pi G \mu a v_\chi =
       \varphi_\chi \left( \dot \varphi_\chi
       - {3 \over 2} H \varphi_\chi \right),
   \label{varphi_chi-eq} \\
   & & \dot \varphi_v =
       {1 \over 2a} \Delta^{-1} \nabla^\alpha \left(
       v_\chi^{\;\;,\beta} \varphi_{v,\alpha\beta}
       + v_{\chi,\alpha} \Delta \varphi_v \right).
   \label{varphi_v-eq}
\eea Equations (\ref{alpha_v-eq}), (\ref{delta_v-eq}),
(\ref{kappa_v-eq}), and (\ref{kappa_v-v_chi-eq}) follow from Eqs.\
(\ref{scalar-6}), (\ref{scalar-5}), (\ref{scalar-3}), and
(\ref{scalar-2}), respectively, evaluated in the comoving gauge. In
Eq.\ (\ref{kappa_v-v_chi-eq}) we used $\chi_v + a v_\chi =
\chi_v^{(q)} + a v_\chi^{(q)}$ and $\chi_v^{(q)} |_v = 0$; see Sec.
VI.C.2 of \cite{NL}. Equation (\ref{alpha_chi-eq}) follows from Eq.\
(\ref{scalar-4}) evaluated in the zero-shear gauge. Equation
(\ref{Poisson-eq}) follows from Eqs.\ (\ref{scalar-1}),
(\ref{scalar-2}), and using $\delta \mu_v \equiv \delta \mu - \dot
\mu a v + \delta \mu_v^{(q)}$, $\varphi_\chi \equiv \varphi - H \chi
+ \varphi_\chi^{(q)}$ and $\varphi_\chi^{(q)} |_\chi = 0$. Equation
(\ref{v_chi-eq}) follows from Eq.\ (\ref{scalar-6}) evaluated in the
zero-shear gauge. Equation (\ref{varphi_chi-eq}) follows from Eqs.\
(\ref{scalar-0}), (\ref{scalar-2}), removing $\kappa$ term and
evaluating in the zero-shear gauge. Equation (\ref{varphi_v-eq})
follows from Eqs.\ (\ref{scalar-0}), (\ref{scalar-2}), removing
$\kappa$ term and evaluating in the comoving gauge. In this set of
equations we located the pure quadratic terms and the possible
second-order pressure terms on the RHSs.

Our notation with a perturbed-order variable as a subindex, for
example, $\delta_v$ indicates a unique gauge-invariant combination
of $\delta$ and $v$ which becomes $\delta$ under the comoving gauge
condition $v = 0$. Thus, $\delta$ in the comoving gauge is {\it
equivalent} to a unique gauge-invariant combination $\delta_v$. To
the linear order we have $\delta_v \equiv \delta - a (\dot \mu/\mu)
v$. An explicit form of $\delta_v$ to the second order and other
gauge-invariant combinations can be found in Eqs.\ (280)-(284) of
\cite{NL}. As we can construct many (in fact, infinitely many) gauge
invariant combinations for $\delta$, our notation apparently has the
advantage of showing explicitly which gauge-invariant combination we
are considering \cite{Bardeen-GI}.

Here, we briefly discuss a conserved variable to the second order.
{}From Eqs.\ (\ref{varphi_v-eq}), (\ref{v_chi-eq}), and
(\ref{alpha_chi-eq}) we have \bea
   & & {1 \over a^3} \left( a^3 \dot \varphi_v \right)^\cdot
       = - {1 \over 2 a^2} \Delta^{-1} \nabla^\alpha \nabla^\beta
       \left( \varphi_{v,\alpha} \varphi_{v,\beta} \right).
\eea
To the linear order we have
\bea
   & & \varphi_v = C ({\bf x}).
   \label{conservation}
\eea Thus, $\varphi_v$ remains constant in time. In the large-scale
limit (super-horizon scale), ignoring the quadratic-order spatial
gradient terms, Eq.\ (\ref{conservation}) remains valid even to the
second order; for more general proof considering the pressure term
see \cite{Salopek-Bond-1990,NL}.

%
%
\section{Issue of pressure}
                                                    \label{sec:Pressures}

Now, we discuss the role of pressure terms in a medium without
pressure. {}From Eqs.\ (233), (235) of \cite{NL} we notice that the
gauge (coordinate) transformation to the second order causes
pressure (both isotropic and anisotropic) terms to appear even in
the case without pressure originally (physically). Such a
complication occurs because our fluid quantities introduced in
\cite{NL} are based on the normal-frame four-vector $\tilde n_a$
which differs from the fluid four-vector $\tilde u_a$. In \cite{NL}
we have presented the fluid quantities based on $\tilde u_a$
separately as well, see Eqs.\ (87), (88) of \cite{NL}; by using
these equations we can translate fluid quantities in the normal
frame to the ones in the fluid frame, and vice versa; the gauge
transformation properties of the fluid quantities in the fluid frame
are presented in Eq.\ (238) of \cite{NL}. The isotropic and
anisotropic pressures are gauge (coordinate) dependent quantities.
To the linear order in the Friedmann background the anisotropic
pressure is gauge invariant and the perturbed isotropic pressure
depends on the coordinate only if we have nonvanishing (and time
varying) background pressure. In the normal-frame, the pure
coordinate transformation to the second and higher orders will cause
both pressures (i.e., isotropic and anisotropic pressure like terms
in the energy-momentum tensor) generated even in the case of
vanishing pressures to the background and to the linear order, see
Eq.\ (233) of \cite{NL}; the frame dependence of fluid quantities
was studied in \cite{pressure-frame}. This complication does not
occur for the fluid quantities based on the fluid frame-vector
$\tilde u_a$; see Eq.\ (238) in \cite{NL}.

{}For vanishing pressure terms in the background and first-order
perturbations we have the following gauge-invariant combinations of
pressure terms (based on $\tilde n_a$) \cite{p-GI} \bea
   \delta p_v
   &=& \delta p - {1 \over 3} \mu v^{,\alpha} v_{,\alpha},
   \nonumber \\
   \Pi_v
   &=& \Pi
       - {3 \over 2} \mu a^2 \Delta^{-2} \nabla^\alpha \nabla^\beta
       \left( v_{,\alpha} v_{,\beta}
       - {1 \over 3} g^{(3)}_{\alpha\beta} v^{,\gamma} v_{,\gamma} \right).
   \nonumber \\
   \label{p_v-GI}
\eea {}From this we notice that the gauge-invariant combination
$\delta p_v$ is the same as $\delta p$ in the comoving gauge.
Evaluating Eq.\ (\ref{p_v-GI}) in the zero-shear gauge ($\chi \equiv
0$) and using $v_\chi \equiv v - {1 \over a} \chi$ to the linear
order, we have \bea
   \delta p_\chi
   &=& \delta p_v
       + {1 \over 3} \mu v_\chi^{\;\;,\alpha} v_{\chi,\alpha},
   \nonumber \\
   \Pi_\chi
   &=& \Pi_v
   \nonumber \\
   & &
       + {3 \over 2} a^2 \mu \Delta^{-2} \nabla^\alpha \nabla^\beta
       \left( v_{\chi,\alpha} v_{\chi,\beta}
       - {1 \over 3} g^{(3)}_{\alpha\beta}
       v_\chi^{\;\;,\gamma} v_{\chi,\gamma} \right).
   \nonumber \\
\eea
As the definition of fluid without pressure we {\it set} the
pressure terms in the comoving gauge equal to be zero, thus
\bea
   & & \delta p_v \equiv 0 \equiv \Pi_v,
   \label{p_v}
\eea
which are gauge-invariant (and physical) zero-pressure conditions.
Thus,
\bea
   \delta p_\chi
   &=&
       {1 \over 3} \mu v_\chi^{\;\;,\alpha} v_{\chi,\alpha}, \quad
   \nonumber \\
   \Pi_\chi
   &=&
       {3 \over 2} a^2 \mu \Delta^{-2} \nabla^\alpha \nabla^\beta
       \left( v_{\chi,\alpha} v_{\chi,\beta}
       - {1 \over 3} g^{(3)}_{\alpha\beta}
       v_\chi^{\;\;,\gamma} v_{\chi,\gamma} \right).
   \nonumber \\
   \label{p_chi}
\eea We set the pressure terms using Eqs.\ (\ref{p_v}),
(\ref{p_chi}). Thus, for fluid quantities based on the normal-frame,
in the gauge other than the comoving gauge the physical
zero-pressure condition implies presence of pressure terms in the
definition of the energy-momentum tensor.

In the comoving gauge without rotation the two frames, $\tilde u_a$
and $\tilde n_a$, coincide. The normal frame $\tilde n_a$ has
$\tilde n_\alpha \equiv 0$. The fluid quantities are ordinarily
defined in the fluid ($\tilde u_a$) frame which differs in general
from the normal four-vector $\tilde n_a$. In the normal-frame
information about the fluid motion is present in the flux
four-vector $\tilde q_a$ with $\tilde q_a \tilde n^a \equiv 0$. In
the energy frame, which takes vanishing flux $\tilde q_a \equiv 0$
as the frame condition, the comoving gauge condition takes $\tilde
u_\alpha \equiv 0$ for the fluid four-vector; here, we ignore the
vector-type perturbation. Since $\tilde u_\alpha = 0$ it coincides
with the normal frame vector. Now, in the normal frame, which takes
$\tilde n_\alpha \equiv 0$ as the frame condition, the comoving
gauge condition without rotation implies $\tilde q_a \equiv 0$.
Thus, as long as we take the comoving gauge without rotation, in
either frame we have $\tilde q_a \equiv 0$ and $\tilde u_\alpha = 0
= \tilde n_\alpha$; i.e., the fluid four-vector coincides with the
normal four-vector.

%
%
\section{A proof}
                                                \label{sec:Proof}

Now, we come to our main point proving the relativistic -Newtonian
correspondence to the second order. Combining Eqs.\
(\ref{delta_v-eq}), (\ref{kappa_v-eq}) we can derive \cite{NL-eq}
\bea
   \ddot \delta_v
   &+& 2 H \dot \delta_v - 4 \pi G \mu \delta_v
   \nonumber \\
   &=& {1 \over a^2} {\partial \over \partial t}
       \left[ a \left( \delta_v v_\chi^{\;\;,\alpha} \right)_{,\alpha} \right]
       + {\Delta \over 2 a^2}
       \left( v_\chi^{\;\;,\alpha} v_{\chi,\alpha} \right).
   \label{ddot-delta_v-eq}
\eea Equations (\ref{delta_v-eq}), (\ref{kappa_v-eq}),
(\ref{Poisson-eq}), and (\ref{ddot-delta_v-eq}) can be compared with
the Newtonian perturbation equations.

The mass conservation, the momentum conservation, and the Poisson's
equation in Newtonian context give \cite{Peebles} \bea
   \dot \delta + {1 \over a} \nabla \cdot {\bf u}
   &=& - {1 \over a} \nabla \cdot \left( \delta {\bf u} \right),
   \label{dot-delta-eq-N} \\
   \dot {\bf u} + H {\bf u} + {1 \over a} \nabla \delta \Phi
   &=& - {1 \over a} {\bf u} \cdot \nabla {\bf u},
   \label{dot-delta-v-eq-N} \\
   {1\over a^2} \nabla^2 \delta \Phi
   &=& 4 \pi G \varrho \delta.
   \label{perturbed-Poisson-eq-N}
\eea
{}From these we have
\bea
   \ddot \delta
   &+& 2 H \dot \delta - 4 \pi G \varrho \delta
   \nonumber \\
   &=& - {1 \over a^2} {\partial \over \partial t}
       \left[ a \nabla \cdot \left( \delta {\bf u} \right) \right]
       + {1 \over a^2} \nabla \cdot \left( {\bf u} \cdot
       \nabla {\bf u} \right).
   \label{ddot-delta-eq-N}
\eea In the Newtonian context Eqs.\
(\ref{dot-delta-eq-N})-(\ref{ddot-delta-eq-N}) are valid to fully
nonlinear order; i.e., the zero-pressure Newtonian fluid equations
are exact in quadratic order nonlinearity. Equation
(\ref{ddot-delta-eq-N}) has been analysed extensively in the
Newtonian context, see \cite{Peebles-1980,quasilinear}.

To the {\it linear order} it is well known that $\delta_v$, $-
\nabla v_\chi$ and $- \varphi_\chi$ (or $\alpha_\chi$) correspond to
a density perturbation ($\delta \equiv {\delta \varrho \over
\varrho}$ with $\tilde \varrho \equiv \varrho + \delta \varrho$ and
$\tilde \varrho$ the mass density), a velocity perturbation (${\bf
u}$) and a perturbation of the gravitational potential ($\delta
\Phi$), respectively.
\cite{Bardeen-1980,H-MDE-1994,HN-Newtonian-1999}. To the {\it linear
order} we may identify \cite{HN-Newtonian-1999} \bea
   & & \delta = \delta_v, \quad
       \delta \Phi = - \varphi_\chi = \alpha_\chi,
   \nonumber \\
   & & {\bf u} \equiv - \nabla v_\chi, \quad
       - {1 \over a} \nabla \cdot {\bf u}
       = {\Delta \over a} v_\chi
       = \kappa_v.
   \label{identify-linear-order}
\eea
As we {\it identify}
\bea
   & & \delta_v = \delta, \quad
       \kappa_v \equiv - {1 \over a} \nabla \cdot {\bf u},
   \label{identify-second-order}
\eea to the {\it second order}, Eq.\ (\ref{ddot-delta_v-eq})
coincides exactly with Eq.\ (\ref{ddot-delta-eq-N}). Equation
(\ref{delta_v-eq}) becomes \bea
   & & \dot \delta_v + {1 \over a} \nabla \cdot {\bf u}
       = - {1 \over a} \nabla \cdot \left( \delta_v {\bf u} \right),
   \label{dot-delta_v-eq-N}
\eea which coincides with Eq.\ (\ref{dot-delta-eq-N}). Equation
(\ref{kappa_v-eq}) gives \bea
   & & \nabla \cdot \left( \dot {\bf u} + H {\bf u} \right)
       + 4 \pi G \mu a \delta_v
       = - {1 \over a}
       \nabla \cdot \left( {\bf u} \cdot \nabla {\bf u} \right),
   \label{dot-delta-u-eq-N}
\eea which also follows from Eqs.\ (\ref{dot-delta-v-eq-N}),
(\ref{perturbed-Poisson-eq-N}) in the Newtonian context. This
completes our proof of the correspondence. Such identifications of
density and velocity perturbations imply that we {\it cannot}
identify $- \varphi_\chi$ (or $\alpha_\chi$) with $\delta \Phi$ to
the second order. This conclusion follows from a close examination
of Eqs.\ (\ref{alpha_v-eq})-(\ref{varphi_v-eq}). In fact, using the
intrinsic three-space curvature in Eq.\ (265) of \cite{NL} \bea
   & & R^{(h)} = {2 \over a^2} \left[ - 2 \Delta \varphi
       + 8 \varphi \Delta \varphi + 3 \varphi^{,\alpha} \varphi_{,\alpha}
       \right],
   \label{R_h}
\eea Eq.\ (\ref{Poisson-eq}) becomes \bea
   & & 4 \pi G \mu \delta_v - {1 \over 4} R^{(h)}_\chi
       = {1 \over 2} \dot H \Delta v_\chi^2
       - 3 a H \dot H \Delta^{-1} \nabla^\alpha
       \left( \delta_{v,\alpha} v_\chi \right),
   \nonumber \\
   \label{Poisson-eq2}
\eea
which still differs from the Newtonian Poisson's equation.
Thus, we conclude that we do not have a relativistic variable which
corresponds to the Newtonian potential to the second order.
Apparently, it is essentially important to employ mixed gauge conditions,
i.e., take different gauge conditions for different variables,
to make correspondence with the Newtonian system:
in this way, correct identifications of (gauge-invariant) variables are
important to show the relativistic-Newtonian correspondence.

At this point, let us clarify the meaning of the quantities involved
in Eqs.\ (\ref{identify-linear-order}),
(\ref{identify-second-order}). Variables $\alpha$, $\chi$ and
$\varphi$ are defined in the metric in Eq.\ (\ref{metric}).
Variables $\chi$ and $\varphi$ can be further identified as the
perturbed shear and perturbed three-space curvature of the normal
hypersurface, respectively. {}From Eq.\ (\ref{R_h}) we find that the
intrinsic scalar curvature $R^{(h)}$ vanishes for $\varphi = 0$.
{}From Eq.\ (264) of \cite{NL} we find that the tracefree part of
the extrinsic curvature tensor $\bar K_{\alpha\beta}$ (equivalently,
shear tensor of the normal frame vector with a minus sign) vanishes
for $\chi = 0$, The variable $\kappa$ can be interpreted as the
perturbed expansion with a minus sign. {}From Eqs.\ (57), (99), and
(179) of \cite{NL} we have $K = - 3 H + \kappa$ where $K$ is a trace
of the extrinsic curvature tensor $K_{\alpha\beta}$ (equivalently,
the expansion scalar, $\tilde \theta \equiv \tilde n^a_{\;\; ;a}$,
with a minus sign). Variables $\delta$ and $v$ are defined in Eq.\
(\ref{Tab-pert-normal}) and can be interpreted as the perturbed
energy-density ($\delta \equiv {\delta \mu \over \mu}$ with $\tilde
\mu = \mu + \delta \mu$) and the flux of the normal-frame,
respectively. In the normal frame, from Eqs.\ (4), (76), and (175)
of \cite{NL} we have the flux vector becomes $J_\alpha \equiv -
\tilde n_b \tilde T^b_\alpha = - a \mu v_{,\alpha}$.

Here we discuss the relation between the comoving and the
synchronous gauge to the second order. Equation (\ref{alpha_v-eq})
shows that $\alpha_v$, which is the same as $\alpha$ in the comoving
gauge ($v \equiv 0$), does not vanish to the second order. This
means that the comoving gauge does not imply our synchronous gauge
to the second order in a zero-pressure medium. At this point it is
important to remember that we already have fixed the spatial gauge
condition using $\gamma \equiv 0$. The original synchronous gauge
used by Lifshitz fixes $\delta g_{00} \equiv 0 \equiv \delta
g_{0\alpha}$, thus $\alpha \equiv 0$ for the temporal gauge and
$\beta \equiv 0$ for the spatial gauge condition. We prefer to fix
$\gamma \equiv 0$ (spatial $C$-gauge) as the spatial gauge condition
instead of $\beta \equiv 0$ (spatial $B$-gauge) because the latter
condition fails to fix the spatial gauge degree of freedom
completely whereas the first one fixes it completely; this is true
even to the second order, and in fact to all orders, in
perturbations, see \S VI.B.2 and VI.C of \cite{NL}. We can show that
the comoving temporal gauge ($v \equiv 0$) together with spatial
$B$-gauge ($\beta = 0$) implies $\alpha = 0$ even to the second
order, for a proof see \cite{CG-SG}. By imposing the comoving ($v
\equiv 0$) and the synchronous ($\alpha \equiv 0$) gauge conditions
simultaneously, Kasai \cite{Kasai-1992} has derived a different
equation compared with ours: such a redundant choice is allowed as
one takes $\beta = 0$ as the spatial gauge condition. However, in
that gauge condition (the spatial $B$-gauge) the spatial gauge-mode
is incompletely fixed, and the comparison with the Newtonian
analyses is {\it not} available.

%
%
\section{Fully nonlinear equations}
                                                     \label{sec:NL}

By extending our comoving gauge condition to be valid to all orders
we can formally derive the {\it completely nonlinear} equations for
the density and velocity perturbations. We will present two methods
to reach such nonlinear equations. These are based on the ADM
($3+1$) equations and the covariant ($1+3$) equations summarised in
\S II.A and II.B, respectively, of \cite{NL}. With the hindsight
gained from our second-order perturbations in previous sections, it
is best to take the comoving gauge condition to all orders. In the
normal-frame context, only the comoving gauge allows the
zero-pressure conditions to be, by definition, vanishing pressure
terms to all orders. To the second order, all the equations we need
to derive Eqs.\ (\ref{ddot-delta_v-eq}), (\ref{dot-delta_v-eq-N}),
and (\ref{dot-delta-u-eq-N}) are Eqs.\ (\ref{alpha_v-eq}),
(\ref{delta_v-eq}), and (\ref{kappa_v-eq}) which follow from Eqs.\
(\ref{scalar-3}), (\ref{scalar-5}), and (\ref{scalar-6}); these are
the Raychaudhury, the energy conservation and the momentum
conservation equations, respectively. We have presented a redundant
set of equations in (\ref{alpha_v-eq})-(\ref{varphi_v-eq}) in order
to show the relativistic-Newtonian correspondences with some
perspective.

The complete set of ADM ($3+1$) equations is presented in Eqs.\
(8)-(13) of \cite{NL}, see \cite{ADM} for original studies. We only
need Eqs.\ (10), (12), and (13) of \cite{NL} which are the trace of
ADM propagation equation, and the energy and momentum conservation
equations, respectively. We take the comoving gauge condition to all
orders which makes the flux four-vector to vanish, i.e., $J_\alpha
\equiv 0$; here we {\it assume} vanishing vector-type perturbation,
thus irrotational, which could contribute to $J_\alpha$. Under such
conditions the zero-pressure conditions (in our normal frame) imply
$S \equiv 0 \equiv \bar S_{\alpha\beta}$ to all orders; $S$ and
$\bar S_{\alpha\beta}$ are the trace and tracefree parts,
respectively, of the spatial part of energy-momentum tensor.
Equation (13) of \cite{NL} gives \bea
   & & N_{,\alpha} = 0,
\eea where $N$ is defined as $\tilde g^{00} \equiv - N^{-2}$. Thus,
we may set $N \equiv a(t)$ to all orders. In this case we have, for
example, $\dot E \equiv E_{,0} N^{-1}$. Now, Eqs.\ (12), (10) of
\cite{NL} become \bea
   \hat {\dot E} - K E
   &=& 0,
   \label{ADM-eq1} \\
   \hat {\dot K}
       - { 1\over 3} K^2
       - \bar K^{\alpha\beta} \bar K_{\alpha\beta}
       - 4 \pi G E + \Lambda
   &=& 0,
   \label{ADM-eq2}
\eea where $\hat {\dot E} \equiv \dot E - E_{,\alpha} N^\alpha
N^{-1}$, {\it etc.}; $E$ is the energy density based on normal frame
vector, and $K$ and $\bar K_{\alpha\beta}$ are the trace and
tracefree parts, respectively, of the extrinsic curvature;
$N_\alpha$ is defined as $\tilde g_{0\alpha} \equiv N_\alpha$. The
spatial indices in ADM formulation are based on the spatial metric
$h_{\alpha\beta}$ defined as $h_{\alpha\beta} \equiv \tilde
g_{\alpha\beta}$. By combining these equations we have \bea
   & & \left( {\hat {\dot E} \over E} \right)^{\hat \cdot}
       - {1 \over 3} \left( {\hat {\dot E} \over E} \right)^2
       - \bar K^{\alpha\beta} \bar K_{\alpha\beta}
       - 4 \pi G E + \Lambda
       = 0.
   \label{ADM-eq3}
\eea Notice again that Eqs.\ (\ref{ADM-eq1})-(\ref{ADM-eq3}) are
valid to all orders, i.e., these equations are fully nonlinear.
{}From Eqs.\ (\ref{ADM-eq1})-(\ref{ADM-eq3}), using \bea
   & & E \equiv \mu + \delta \mu,
\eea and the quantities presented in \cite{NL} we can easily derive
Eqs.\ (\ref{dot-delta_v-eq-N}), (\ref{dot-delta-u-eq-N}), and
(\ref{ddot-delta_v-eq}), respectively; see the next section.

The complete set of covariant ($1+3$) equations is presented in
Eqs.\ (26)-(37) of \cite{NL}; see \cite{covariant} for original
studies. We only need Eqs.\ (26)-(28) of \cite{NL} which are the
energy and momentum conservations and the Raychaudhury equation,
respectively. We take the energy-frame which sets the energy flux
term to vanish, i.e., $\tilde q_a \equiv 0$. In this frame the frame
four-vector $\tilde u_a$ is the fluid four-vector. The zero-pressure
conditions imply $\tilde p \equiv 0 \equiv \tilde \pi_{ab}$ to all
orders; $\tilde \pi_{ab}$ is the covariant anisotropic stress based
on $\tilde u_a$. Equation (27) of \cite{NL} gives vanishing
acceleration vector, i.e., $\tilde a_a \equiv \tilde u_{a;b} \tilde
u^b = 0$ to all orders. Thus, Eqs.\ (26), (28) of \cite{NL} become
\bea
   \tilde {\dot {\tilde \mu}} + \tilde \mu \tilde \theta
   &=& 0,
   \label{covariant-eq1} \\
   \tilde {\dot {\tilde \theta}} + {1 \over 3} \tilde \theta^2
       + \tilde \sigma^{ab} \tilde \sigma_{ab}
       - \tilde \omega^{ab} \tilde \omega_{ab}
       + 4 \pi G \tilde \mu - \Lambda
   &=& 0,
   \label{covariant-eq2}
\eea where $\tilde \theta \equiv \tilde u^a_{\;\; ;a}$ is an
expansion scalar, and $\tilde \sigma_{ab}$ is the shear tensor. An
overdot with tilde is a covariant derivative along the $\tilde u_a$
vector, e.g., $\tilde {\dot {\tilde \mu}} \equiv \tilde \mu_{,a}
\tilde u^a$. By combining these equations we have \bea
   & & \left( {\tilde {\dot {\tilde \mu}} \over \tilde \mu}
       \right)^{\tilde \cdot}
       - {1 \over 3}
       \left( {\tilde {\dot {\tilde \mu}} \over \tilde \mu} \right)^2
       - \tilde \sigma^{ab} \tilde \sigma_{ab}
       + \tilde \omega^{ab} \tilde \omega_{ab}
       - 4 \pi G \tilde \mu
       + \Lambda
       = 0.
   \nonumber \\
   \label{covariant-eq3}
\eea Notice that Eqs.\ (\ref{covariant-eq1})-(\ref{covariant-eq3})
are valid to all orders, i.e., these equations are fully nonlinear.
More general equation in a fully covariant form considering the
general pressure terms can be found in Eq.\ (88) of
\cite{HV-covariant-1990}.

We take the comoving gauge condition to all orders which makes the
space part of four-vector with low index to vanish, i.e., $\tilde
u_\alpha \equiv 0$; here we also {\it assume} vanishing vector-type
perturbation, thus irrotational, which could contribute to $\tilde
u_\alpha$. As our gauge condition (and the irrotational condition)
implies $\tilde u_\alpha \equiv 0$, the frame vector is the same as
the normal frame, thus $\tilde u_a = \tilde n_a$. In such a case we
have vanishing rotation of the $\tilde u_a$ flow, thus $\tilde
\omega_{ab} = 0$. {}From Eqs.\
(\ref{covariant-eq1})-(\ref{covariant-eq3}), using \bea
   & & \tilde \mu \equiv \mu + \delta \mu,
\eea and the quantities presented in \cite{NL} we can easily derive
Eqs.\ (\ref{dot-delta_v-eq-N}), (\ref{dot-delta-u-eq-N}), and
(\ref{ddot-delta_v-eq}), respectively. A derivation based on the
covariant equations is presented in \cite{Second-CQG}.

Afterall, the ADM equations (\ref{ADM-eq1})-(\ref{ADM-eq3}) are the
same as the covariant equations
(\ref{covariant-eq1})-(\ref{covariant-eq3}), expressed in different
forms. We can derive the ADM equations by rewriting the covariant
equations in the normal frame vector. Since our comoving gauge
condition with irrotational condition makes $\tilde u_\alpha \equiv
0$, the frame vector is the same as the normal frame vector $\tilde
n_a$. By direct calculations, using the quantities presented in
Eqs.\ (2)-(6), and (14)-(16) of \cite{NL}, we can show that \bea
   & & \tilde \mu = E, \quad
       \tilde \theta = - K, \quad
       \tilde \sigma^{ab} \tilde \sigma_{ab}
       = \bar K^{\alpha\beta} \bar K_{\alpha\beta},
   \nonumber \\
   & & \tilde {\dot {\tilde \mu}} = \hat {\dot E}, \quad
       \tilde {\dot {\tilde \theta}} = - \hat {\dot K}.
\eea Using this we can show that Eqs.\
(\ref{covariant-eq1})-(\ref{covariant-eq3}) give Eqs.\
(\ref{ADM-eq1})-(\ref{ADM-eq3}); these equations are valid
considering general background curvature and the tensor-type
perturbation (gravitational waves) to all orders.

%
%
\section{Another derivation including the gravitational waves}
                                                     \label{sec:Proof2}

Since Eqs.\ (\ref{dot-delta_v-eq-N}), (\ref{dot-delta-u-eq-N}) are
our main results allowing us to conclude about the
relativistic-Newtonian correspondence, in the following we will
derive these equations in some detail again directly from the fully
nonlinear equations in \S \ref{sec:NL}. Now, we include the
gravitational wave contribution. The metric becomes \bea
   ds^2
   &=& - a^2 \left( 1 + 2 \alpha \right) d \eta^2
       - 2 a \chi_{,\alpha} d \eta d x^\alpha
   \nonumber \\
   & &
       + a^2 \left[ \left( 1 + 2 \varphi \right) \delta_{\alpha\beta}
       + 2 C^{(t)}_{\alpha\beta} \right] d x^\alpha d x^\beta,
   \label{metric-2}
\eea where $C^{(t)}_{\alpha\beta}$ is the transverse and tracefree
gravitational waves; its indices are based on
$g^{(3)}_{\alpha\beta}$. We work in the temporal comoving gauge.
Thus, $C^{(t)}_{\alpha\beta}$ is also evaluated in the comoving
gauge, and equivalent to a gauge-invariant combination
$C^{(t)}_{\alpha\beta v}$.

We introduce \bea
   & & E \equiv \mu + \delta \mu, \quad
       K \equiv - 3 {\dot a \over a} + \kappa,
\eea see Eqs.\ (45), (72), (178), and (179) of \cite{NL}. We have
\bea
   & & \hat {\dot E}
       \equiv \dot E - E_{,\alpha} N^\alpha N^{-1}
       = \dot \mu + \delta \dot \mu
       + {1 \over a^2} \delta \mu_{,\alpha} \chi^{,\alpha},
   \nonumber \\
   & & \hat {\dot K}
       \equiv \dot K - K_{,\alpha} N^\alpha N^{-1}
       = - 3 \left( {\dot a \over a} \right)^\cdot
       + \dot \kappa
       + {1 \over a^2} \kappa_{,\alpha} \chi^{,\alpha}.
\eea In setting $N = a$ we already have used the comoving gauge
condition. Since we take the comoving gauge we often ignore the
subindex $v$ which indicates the comoving gauge choice (equivalently
the unique corresponding gauge-invariant combination between the
variable and $v$); for example, our $\delta$ is the same as a
gauge-invariant combination $\delta_v$ which is the same as $\delta$
in the comoving gauge setting $v \equiv 0$. Using Eqs.\ (55), (57),
and (175) of \cite{NL} we can show \bea
   \bar K^{\alpha\beta} \bar K_{\alpha\beta}
   &=& {1 \over a^4} \left[ \chi^{\;,\alpha|\beta} \chi_{,\alpha|\beta}
       - {1 \over 3} \left( \Delta \chi \right)^2 \right]
   \nonumber \\
   & &
       + \dot C^{(t)\alpha\beta} \left( {2 \over a^2} \chi_{,\alpha|\beta}
       + \dot C^{(t)}_{\alpha\beta} \right).
\eea Equations (\ref{ADM-eq1}), (\ref{ADM-eq2}) become \bea
   & & \left( {\dot \mu \over \mu} + 3 {\dot a \over a} \right)
       \left( 1 + \delta \right)
       + \dot \delta - \kappa
       = \kappa \delta
       - {1 \over a^2} \delta_{,\alpha} \chi^{,\alpha},
   \\
   & & 3 {\ddot a \over a} + 4 \pi G \mu - \Lambda
       - \dot \kappa - 2 {\dot a \over a} \kappa
       + 4 \pi G \mu \delta
   \nonumber \\
   & & \quad
       = {1 \over a^2} \kappa_{,\alpha} \chi^{,\alpha}
       - { 1\over 3} \kappa^2
       - {1 \over a^4} \left[ \chi^{,\alpha|\beta}
       \chi_{,\alpha|\beta}
       - {1 \over 3} \left( \Delta \chi \right)^2 \right]
   \nonumber \\
   & & \quad
       - \dot C^{(t)\alpha\beta} \left( {2 \over a^2} \chi_{,\alpha|\beta}
       + \dot C^{(t)}_{\alpha\beta} \right).
\eea

Now, we have to relate $\chi (\equiv \chi_v)$ to our notation.
Apparently, we need $\chi$ only to the linear order. To the linear
order the $\tilde G^0_\alpha$-component of Einstein equation in Eq.\
(\ref{kappa_v-v_chi-eq}) gives ${\Delta \over a^2} \chi_v = -
\kappa_v \equiv {1 \over a} \nabla \cdot {\bf u}$; we have $\chi_v
\equiv \chi - a v \equiv - a v_\chi$ to the linear order. As our
${\bf u}$ is of the potential type, i.e., of the form ${\bf u}
\equiv u_{,\alpha}$, we have \bea
   & & {\bf u} = {1 \over a} \nabla \chi_v,
\eea to the linear order. Thus, we have \bea
   & & \left( {\dot \mu \over \mu} + 3 {\dot a \over a} \right)
       \left( 1 + \delta_v \right)
       + \dot \delta_v + {1 \over a} \nabla \cdot {\bf u}
       = - {1 \over a} \nabla \cdot \left( \delta_v {\bf u} \right),
   \\
   & & 3 {\ddot a \over a} + 4 \pi G \mu - \Lambda
       + {1 \over a} \nabla \cdot \left( \dot {\bf u}
       + {\dot a \over a} {\bf u} \right)
       + 4 \pi G \mu \delta_v
   \nonumber \\
   & & \quad
       = - {1 \over a^2}
       \nabla \left( {\bf u} \cdot \nabla {\bf u} \right)
       - \dot C^{(t)\alpha\beta} \left( {2 \over a} u_{\alpha,\beta}
       + \dot C^{(t)}_{\alpha\beta} \right).
\eea The perturbed parts give Eqs.\ (\ref{dot-delta_v-eq-N}),
(\ref{dot-delta-u-eq-N}) with additional contributions from the
gravitational waves in Eq.\ (\ref{dot-delta-u-eq-N}), thus in Eq.\
(\ref{ddot-delta_v-eq}) as well.

Therefore, in the presence of the tensor-type perturbation we have
\bea
   & & \dot \delta_v + {1 \over a} \nabla \cdot {\bf u}
       = - {1 \over a} \nabla \cdot \left( \delta_v {\bf u} \right),
   \label{dot-delta-eq} \\
   & & {1 \over a} \nabla \cdot \left( \dot {\bf u}
       + {\dot a \over a} {\bf u} \right)
       + 4 \pi G \mu \delta_v
        = - {1 \over a^2} \nabla \cdot
       \left( {\bf u} \cdot \nabla {\bf u} \right)
   \nonumber \\
   & & \quad
       - \dot C^{(t)\alpha\beta} \left( {2 \over a} u_{\alpha,\beta}
       + \dot C^{(t)}_{\alpha\beta} \right),
   \label{dot-delta-u-eq}
\eea
thus
\bea
   & & \ddot \delta_v
       + 2 {\dot a \over a} \dot \delta_v
       - 4 \pi G \mu \delta_v
       = - {1 \over a^2} {\partial \over \partial t}
       \left[ a \nabla \cdot \left( \delta_v {\bf u} \right) \right]
   \nonumber \\
   & & \quad
       + {1 \over a^2} \nabla \cdot \left( {\bf u} \cdot
       \nabla {\bf u} \right)
       + \dot C^{(t)\alpha\beta} \left( {2 \over a} u_{\alpha,\beta}
       + \dot C^{(t)}_{\alpha\beta} \right).
   \label{ddot-delta-eq}
\eea
The presence of linear-order gravitational waves can generate the
second-order scalar-type perturbation by generating the shear terms.
Here, we note the behaviour of the gravitational waves in the linear regime.
To the linear order the gravitational waves are described by the well known
wave equation \cite{Lifshitz-1946}
\bea
   & & \ddot C^{(t)}_{\alpha\beta}
       + 3 {\dot a \over a} \dot C^{(t)}_{\alpha\beta}
       - {\Delta \over a^2} C^{(t)}_{\alpha\beta} = 0.
\eea In the super-horizon scale the non-transient mode of
$C^{(t)}_{\alpha\beta}$ remains constant, thus $\dot
C^{(t)}_{\alpha\beta} = 0$, and in the sub-horizon scale, the
oscillatory $C^{(t)}_{\alpha\beta}$ redshifts away, thus
$C^{(t)}_{\alpha\beta} \propto a^{-1}$. Thus, we anticipate that the
contribution of gravitational waves to the scalar-type perturbation
is not significant to the second order.

To the second order the equation for tensor-type perturbation
(gravitational waves) can be derived from Eqs.\ (103), (210) of
\cite{NL}. Since we are ignoring the vector-type perturbation from
Eqs.\ (211), (199) of \cite{NL} we have \bea
    & & \ddot C^{({t})}_{\alpha\beta}
        + 3 {\dot a \over a} \dot C^{({t})}_{\alpha\beta}
        - {\Delta \over a^2} C^{({t})}_{\alpha\beta}
        = N_{4\alpha\beta}
    \nonumber \\
    & & \quad
        - {3 \over 2} \left( \nabla_\alpha \nabla_\beta
        - {1 \over 3} g^{(3)}_{\alpha\beta} \Delta \right)
       \Delta^{-2} \nabla^{\gamma} \nabla^\delta
       N_{4\gamma\delta},
    \label{GW-eq}
\eea where we assumed a flat background and set anisotropic stress
to be zero. {}From Eq.\ (103) of \cite{NL} to the second order we
have \bea
   & & N_{4 \alpha\beta}
       = {1 \over a^3} \Bigg\{ a^3 \Bigg[
       {2 \over a^2} \left( \varphi \chi_{,\alpha|\beta}
       + \varphi_{,(\alpha} \chi_{,\beta)} \right)
       + 2 \varphi \dot C^{(t)}_{\alpha\beta}
   \nonumber \\
   & & \quad
       + {2 \over a^2} \chi^{,\gamma}_{\;\;\;|\beta} C^{(t)}_{\alpha\gamma}
       + {1 \over a^2} \chi^{,\gamma}
       \left( 2 C^{(t)}_{\gamma(\alpha|\beta)}
       - C^{(t)}_{\alpha\beta|\gamma} \right)
   \nonumber \\
   & & \quad
       + 2 C^{(t)\gamma}_{\;\;\;\;\;\alpha} \dot C^{(t)}_{\beta\gamma}
       \Bigg] \Bigg\}^\cdot
   \nonumber \\
   & & \quad
       + {1 \over a^4} \chi^{,\gamma}_{\;\;\;|\alpha} \chi_{,\gamma|\beta}
       + {1 \over a^2} \left( \kappa \chi_{,\alpha|\beta}
       - 4 \varphi \varphi_{,\alpha|\beta}
       - 3 \varphi_{,\alpha} \varphi_{,\beta} \right)
   \nonumber \\
   & & \quad
       + \kappa \dot C^{(t)}_{\alpha\beta}
       + {1 \over a^2} \Bigg[
       2 \varphi^{,\gamma}_{\;\;\;|\alpha} C^{(t)}_{\beta\gamma}
       - 2 \Delta \varphi C^{(t)}_{\alpha\beta}
       - 4 \varphi \Delta C^{(t)}_{\alpha\beta}
   \nonumber \\
   & & \quad
       + \varphi^{,\gamma} \left(
       2 C^{(t)}_{\gamma(\alpha|\beta)}
       - 3 C^{(t)}_{\alpha\beta|\gamma} \right)
       + 2 \chi^{,\gamma}_{\;\;\;|[\alpha} \dot C^{(t)}_{\beta]\gamma}
       - \chi^{,\gamma} \dot C^{(t)}_{\alpha\beta|\gamma}
   \nonumber \\
   & & \quad
       + 2 C^{(t)\gamma\delta} \left(
       2 C^{(t)}_{\gamma(\alpha|\beta)\delta}
       - C^{(t)}_{\alpha\beta|\gamma\delta}
       - C^{(t)}_{\gamma\delta|\alpha\beta} \right)
   \nonumber \\
   & & \quad
       - 2 C^{(t)\gamma}_{\;\;\;\;\;\alpha} \Delta C^{(t)}_{\beta\gamma}
       - C^{(t)\gamma}_{\;\;\;\;\;\delta|\alpha}
       C^{(t)\delta}_{\;\;\;\;\;\gamma|\beta}
       + 4 C^{(t)\gamma|\delta}_{\;\;\;\;\;\alpha}
       C^{(t)}_{\beta[\delta|\gamma]} \Bigg]
   \nonumber \\
   & & \quad
       - {1 \over 3} g^{(3)}_{\alpha\beta} \Bigg\{
       {1 \over a^3} \Big\{ a^3 \Big[
       {2 \over a^2} \left( \varphi \Delta \chi
       + \varphi^{,\gamma} \chi_{,\gamma} \right)
   \nonumber \\
   & & \quad
       + 2 C^{(t)\gamma\delta} \Big( {1 \over a^2} \chi_{,\gamma|\delta}
       + \dot C^{(t)}_{\gamma\delta} \Big) \Big] \Big\}^\cdot
       + {1 \over a^4} \chi^{,\gamma|\delta} \chi_{,\gamma|\delta}
   \nonumber \\
   & & \quad
       + {1 \over a^2} \Big[ \kappa \Delta \chi
       - 4 \varphi \Delta \varphi
       - 3 \varphi^{,\gamma} \varphi_{,\gamma}
       + 2 \varphi^{,\gamma|\delta} C^{(t)}_{\gamma\delta}
   \nonumber \\
   & & \quad
       - 4 C^{(t)\gamma\delta} \Delta C^{(t)}_{\gamma\delta}
       + C^{(t)\gamma\delta|\epsilon}
       \left( 2 C^{(t)}_{\gamma\epsilon|\delta}
       - 3 C^{(t)}_{\gamma\delta|\epsilon} \right) \Big]
       \Bigg\}.
   \label{N_4}
\eea In Eq.\ (\ref{N_4}) we have ignored $\alpha$ and $\dot \varphi$
terms which are already quadratic order in the comoving gauge, see
Eqs.\ (\ref{alpha_v-eq}), (\ref{varphi_v-eq}). Since we are in the
comoving gauge, we have $\chi = \chi_v$, $\varphi = \varphi_v$,
$\kappa = \kappa_v$ and $C^{(t)}_{\alpha\beta} =
C^{(t)}_{\alpha\beta v}$. Apparently, we need $\chi_v$, $\kappa_v$
and $\varphi_v$ to the linear order. We have $\kappa_v = - {1 \over
a} \nabla \cdot {\bf u}$ and ${\bf u} = {1 \over a} \nabla \chi_v$.
{}For $\varphi_v$ we have \bea
   & & \varphi_v \equiv \varphi - a H v
       = \varphi_\chi - a H v_\chi,
\eea where we have $\varphi_\chi = - \delta \Phi$ and ${\bf u} = -
\nabla v_\chi$ in Eq.\ (\ref{identify-linear-order}). Using these
identifications we can express the scalar-type perturbation
variables in Eq.\ (\ref{N_4}) in terms of the Newtonian variables.

Equations (\ref{dot-delta-eq}), (\ref{dot-delta-u-eq}), and
(\ref{GW-eq}) provide a {\it complete} set describing the scalar-
and tensor-type perturbations to the second order in the flat
Friedmann background. {}From these equations we can see that the
linear-order scalar-type (tensor-type) perturbation works as a
source for the tensor-type (scalar-type) perturbation to the second
order. Such effects and the presence of the gravitational waves are
purely general relativistic ones.

%
%
\section{Discussion}

We have shown that to the second order, ignoring the gravitational
wave contribution, the zero-pressure relativistic cosmological
perturbation equations can be exactly identified with the known
equations in Newtonian system, compare Eqs.\
(\ref{dot-delta-eq})-(\ref{ddot-delta-eq}) with Eqs.\
(\ref{dot-delta-eq-N})-(\ref{ddot-delta-eq-N}). More precisely, the
relativistic equations can be identified with the continuity
equation and the divergence of the Euler equation replacing the
Newtonian gravitational potential using Poisson's equation. In order
to achieve such a correspondence we need correct identification of
gauge-invariant density and velocity perturbation variables as in
Eqs.\ (\ref{identify-linear-order}), (\ref{identify-second-order}).
It is important to notice that we have {\it avoided} using the
potential-like variable in our identification. In fact, we showed
that we do {\it not} have a relativistic variable which corresponds
to the Newtonian potential to the second order. This is
understandable because the gravitational potential introduced in
Poisson's equation reveals the action-at-a-distance nature and the
static nature of Newton's gravity theory compared with the
relativistic gravity.

As a consequence, to the second order, the Newtonian hydrodynamic
equations (\ref{ddot-delta-eq-N}), (\ref{dot-delta_v-eq-N}), and
(\ref{dot-delta-u-eq-N}) remain valid in {\it all} cosmological
scales including the super-horizon scale. Although showing the
equivalence of the zero-pressure relativistic scalar-type
perturbation to the Newtonian ones to the second order, may not be
entirely surprising it should not be so obvious either. It might be
as well that our relativistic results give relativistic correction
terms appearing to the second order which become important as we
approach and go beyond the horizon scale. Our results show that
there are {\it no} such correction terms appearing to the second
order, and the correspondence is exact to that order. A
complementary result, showing the relativistic-Newtonian
correspondence in the Newtonian limit of the post-Newtonian
approach, can be found in \cite{Kofman-Pogosyan-1995}. In fact, the
Newtonian hydrodynamic equations appear naturally as the
zeroth-order post-Newtonian limit \cite{PN}.

We note that although we assumed a flat background, our equations
are valid with the cosmological constant. Thus, these are compatible
with current observations of the large-scale structure and the
cosmic microwave background radiation which favour near flat
Friedmann world model with non-vanishing $\Lambda$
\cite{observations}. As we consider a flat background the ordinary
Fourier analysis can be used to study the mode-couplings as in the
Newtonian case in \cite{quasilinear}. Our result also may have the
following important practical cosmological implication. As we have
proved that the Newtonian hydrodynamic equations are valid in {\it
all} cosmological scales to the second order, our result has an
important cosmological implication that large-scale Newtonian
numerical simulation can be used more reliably in the general
relativistic context even as the simulation scale approaches near
(and goes beyond) the horizon scale.

At this point, it is important to be reminded that we showed the
relativistic-Newtonian correspondence for the density and velocity
perturbations, but not for the gravitational potential. Therefore,
although our result assures that one can trust cold dark matter
simulations  at {\it all} scales for the density and velocity
fields, it does {\it not} imply that one can trust the Newtonian
simulations for effects involving the gravitational potential, like
the weak gravitational lensing effects. Indeed, in order to handle
the lensing effects properly we often require an extra factor of two
which comes from the post-Newtonian effects.\footnote{We thank the
anonymous referee for suggesting this point.}

Since the Newtonian system is exact to the second order in
nonlinearity, besides the gravitational wave contribution to the
second and higher order, any nonvanishing third and higher order
perturbation terms in the relativistic analysis can be regarded as
the pure relativistic corrections. Expanding the fully nonlinear
equations in (\ref{covariant-eq1})-(\ref{covariant-eq3}) or
(\ref{ADM-eq1})-(\ref{ADM-eq3}) to third and higher order will give
the potential correction terms. Our recent investigation of this
important open question shows that to the third order there occur
pure relativistic correction terms which are of $\varphi_v$-order
higher \cite{Third}. Thus, the corrections are independent of the
horizon and are small; see the accompanying contribution in
\cite{Third}.

In this work we have considered an irrotational single component
dust in the flat background. Extending any of these assumptions
could lead to situations which deserve further attention. {}First,
it would be interesting to see up to what point the correspondence
between the two theories can be extended in the non-flat case. In
this way we can identify possible relativistic effects caused by the
non-flat nature of the background. Second, in this work we have
ignored the vector-type perturbation because it simply decays in the
expanding phase. This has to do with considering only the
longitudinal  part of ${\bf u}$ in Eqs.\ (\ref{dot-delta-u-eq-N}),
(\ref{dot-delta-u-eq}). It would be interesting to include the
rotational mode to see the similarity and difference between the two
gravity theories. As the realistic Newtonian simulations include the
whole ${\bf u}$ vector as the perturbed velocity it would be
practically important to see the role of relativistic vector-type
perturbation to the second order, and to determine whether the
relativistic effect could be important. Third, the usual
cosmological simulations include the cold dark matter together with
the baryon, thus a system with two components. Thus, the
relativistic nonlinear perturbations of the zero-pressure but
multi-component system would be interesting subject in practice. It
is, a priori, unclear whether the relativistic-Newtonian
correspondence would continue in such a multi-component case. In the
second and the third subjects, the comoving gauge issue should be
applied with care. {}Fourth, the presence of substantial amount of
pressures (both isotropic and anisotropic) would lead to
relativistic corrections. Even in the linear perturbation the
pressure terms cause new relativistic correction terms which are not
present in the Newtonian system. Thus, including the pressure terms
in relativistic second-order perturbation is interesting because
most of the terms will be pure relativistic corrections. Such a
formulation would be practically interesting because we anticipate
presence of strong pressure in the early universe. All these four
subjects are left for future studies.

%
%
\subsection*{Acknowledgments}

We thank Dirk Puetzfeld for useful suggestions, and Misao Sasaki for
useful discussions on the synchronous gauge issue. HN and JH were
supported by grants No. R04-2003-10004-0 and No.
R02-2003-000-10051-0, respectively, from the Basic Research Program
of the Korea Science and Engineering Foundation.



\begin{thebibliography}{99}
\bibitem{Friedmann-1922}
         A.A. Friedmann, Zeitschrift f\"ur Physik {\bf 10}, 377 (1922),
                         and
                         {\it ibid.} {\bf 21}, 326 (1924);
                         both papers are translated in
                         {\it Cosmological-constants: papers in modern
                         cosmology}, edited by J. Bernstein and G. Feinberg
                         (Columbia Univ. Press, New York, 1986), p49 and p59;
         H.P. Robertson, Proceedings of the National Academy of Science
                         {\bf 15}, 822 (1929).
\bibitem{Lifshitz-1946}
         E.M. Lifshitz, J. Phys. (USSR) {\bf 10}, 116 (1946);
         E.M. Lifshitz and I.M. Khalatnikov, Adv. Phys. {\bf 12}, 185 (1963).
\bibitem{Milne-1934}
         E.A. Milne, Quart. J. Math. {\bf 5}, 64 (1934).
\bibitem{Bonnor-1957}
         W.B. Bonnor, Mon. Not. R. Astron. Soc. {\bf 117}, 104 (1957).
\bibitem{SW-1967}
         See the last paragraph in the Appendix II of
         R.K. Sachs and A.M. Wolfe, Astrophys. J. {\bf 147}, 73 (1967).
\bibitem{Harrison-Nariai}
         E.R. Harrison, Rev. Mod. Phys. {\bf 39}, 862 (1967);
         G.B. Field and L.C. Shepley, Astrophys. Space. Sci.
                    {\bf 1}, 309 (1968);
         H. Nariai, Prog. Theor. Phys. {\bf 41}, 686 (1969);
         V.N. Lukash, Sov. Phys. JETP {\bf 52}, 807 (1980);
         H. Kodama and M. Sasaki,
                   Prog. Theor. Phys. Suppl. {\bf 78}, 1 (1984).
\bibitem{Bardeen-1980}
         J.M. Bardeen, Phys. Rev. D {\bf 22}, 1882 (1980).
\bibitem{GRG}
         See \S 4 in J. Hwang, Gen. Rel. Grav. {\bf 23}, 235 (1991),
         and \S 5 of Ref. \cite{HN-Newtonian-1999}.
\bibitem{HN-Newtonian-1999}
         J. Hwang and H. Noh, Gen. Rel. Grav. {\bf 31}, 1131 (1999).
\bibitem{Bardeen-1988}
         J.M. Bardeen, {\it Particle Physics and Cosmology}, edited by
                       L. Fang and A. Zee
                       (Gordon and Breach, London, 1988), p1;
         J. Hwang, Astrophys. J. {\bf 375}, 443 (1991).
\bibitem{NL}
         H. Noh and J. Hwang, Phys. Rev. D {\bf 69}, 104011 (2004).
\bibitem{H-MDE-1994}
         J. Hwang, Astrophys. J. {\bf 427}, 533 (1994).
\bibitem{MM-1934}
         W.H. McCrea and E.A. Milne, Quart. J. Math. {\bf 5}, 73 (1934).
\bibitem{Bardeen-GI}
         In Bardeen's notation of \cite{Bardeen-1980},
         to the linear order we have:
         $\delta_v \equiv \epsilon_m$,
         $v_\chi \equiv v_s/k$,
         $\varphi_\chi \equiv \Phi_H$,
         $\alpha_\chi \equiv \Phi_A$, and
         $\varphi_v \equiv \phi_m$.
\bibitem{Salopek-Bond-1990}
         D.S. Salopek and J.R. Bond, Phys. Rev. D {\bf 42}, 3936 (1990).
\bibitem{pressure-frame}
         A.R. King and G.F.R. Ellis,
                       Comm. Math. Phys. {\bf 31}, 209 (1973);
         R. Maartens and F.P. Wolvaardt,
                         Class. Quant. Grav. {\bf 11}, 203 (1994);
         R. Maartens, Phys. Rev. D {\bf 58}, 124006 (1998).
\bibitem{p-GI}
         In order to derive these gauge-invariant combinations, follow
         the method described in \S VI.C.2 of \cite{NL}.
\bibitem{NL-eq}
         This equation was derived in Eq.\ (342) of \cite{NL} without
         noticing the issue of zero-pressure condition
         in the normal-frame choice.
         Since we used only Eqs.\ (\ref{delta_v-eq}), (\ref{kappa_v-eq})
         in \cite{NL}, which are all evaluated in the comoving gauge,
         it does not affect the result.
         In \cite{NL} the Newtonian correspondence to the second order
         was addressed incompletely for which we need to examine the other
         equations in Eqs.\ (\ref{alpha_v-eq})-(\ref{varphi_v-eq}).
         The analyses made in \S VII.C of \cite{NL}, although incomplete,
         remain correct.
\bibitem{Peebles}
         These equations can be found in Eqs.\ (9.17), (7.14) of
         \cite{Peebles-1980};
         for an earlier work but not in cosmological context, see
         C. Hunter, Astrophys. J. {\bf 139}, 570 (1964).
\bibitem{Peebles-1980}
         P.J.E. Peebles, {\it The large-scale structure of the universe},
                         (Princeton Univ. Press, Princeton, 1980).
\bibitem{quasilinear}
         E.T. Vishniac, Mon. Not. R. Astron. Soc. {\bf 203}, 345 (1983);
         M.H. Goroff, B. Grinstein, S.-J. Rey, and M.B. Wise,
                       Astrophys. J., {\bf 311}, 6 (1986);
         F. Bernardeau, S. Colombi, E. Gaztanaga, and R. Scoccimarro,
                        Phys. Rep. {\bf 367}, 1 (2002).
\bibitem{CG-SG}
         In the normal frame we have the acceleration vector becomes
         $\tilde a_\alpha = ( \alpha - \alpha^2
         + {1 \over 2} \beta^{,\gamma} \beta_{,\gamma} )_{,\alpha}$,
         see Eq.\ (69) of \cite{NL}.
         In that frame the temporal comoving gauge implies
         $\tilde q_\alpha = 0$, see Eqs.\ (72), (175) of \cite{NL}.
         The zero-pressure condition implies the momentum-conservation
         equation, Eq.\ (27) of \cite{NL}, gives $\tilde a_\alpha = 0$.
         Thus, if we take the spatial $B$ gauge, $\beta \equiv 0$,
         we have $\alpha = 0$ which is the temporal synchronous gauge.
\bibitem{Kasai-1992}
         M. Kasai, Phys. Rev. Lett. {\bf 69}, 2330 (1992);
                   Phys. Rev. D {\bf 47}, 3214 (1993).
\bibitem{ADM}
         R. Arnowitt, S. Deser, and C.W. Misner, in {\it Gravitation: an
                      introduction to current research}, edited by  L. Witten
                      (Wiley, New York, 1962) p. 227.
\bibitem{covariant}
         J. Ehlers, Proceedings of the mathematical-natural science of
                    the Mainz academy of science and literature,
                    Nr. {\bf 11}, 792 (1961),
                    translated in Gen. Rel. Grav. {\bf 25}, 1225 (1993);
         G.F.R. Ellis, in {\it General relativity and cosmology,
                       Proceedings of
                       the international summer school of physics Enrico
                       Fermi course 47}, edited by R. K. Sachs (Academic
                       Press, New York, 1971), p104;
                       in {\it Cargese Lectures in Physics}, edited by
                       E. Schatzmann (Gorden and Breach, New York, 1973), p1.
\bibitem{HV-covariant-1990}
         J. Hwang and E.T. Vishniac, Astrophys. J.  {\bf 353}, 1 (1990);
                  for an earlier work, see
         J.C. Jackson, Proc. R. Soc. Lond. A {\bf 328}, 561 (1972).
\bibitem{Second-CQG}
         H. Noh and J. Hwang, Class. Quant. Grav. in press (2005),
                    astro-ph/0412127.
\bibitem{Kofman-Pogosyan-1995}
         E. Bertschinger and A.J.S. Hamilton,
                         Astrophys. J. {\bf 435}, 1 (1994);
         L. Kofman and D. Pogosyan, {\it ibid.} {\bf 442}, 30
                   (1995);
         G.F.R. Ellis and P.K.S. Dunsby, Astrophys. J. {\bf 479}, 97
                      (1997).
\bibitem{PN}
         S. Chandrasekhar, Astrophys. J. {\bf 142}, 1488 (1965);
         J. Hwang, H. Noh, and D. Puetzfeld, Phys. Rev. D submitted,
                   astro-ph/0507085.
\bibitem{observations}
         D.N. Spergel, {\it et al.},
                       Astrophys. J. Suppl. {\bf 148}, 175 (2003);
         M. Tegmark, {\it et al.}, Phys. Rev. D {\bf 69}, 103501 (2004).
\bibitem{Third}
         H. Noh and J. Hwang, Phy. Rev. D in press (2005), gr-qc/0412129.
\end{thebibliography}
\end{document}